\documentclass[10pt]{article}
\newcommand{\fra}[1]{{\color{black}{{#1}}}}

\usepackage{color}
\usepackage{amssymb,upgreek}
\usepackage{mathptmx}
\usepackage{multicol,xcolor}
\usepackage{anyfontsize}
\usepackage{t1enc,color,xcolor,tikz,hyperref}

\usepackage{graphicx}	% Including figure files
\usepackage{amsmath}	% Advanced maths commands
\usepackage{amssymb}	% Extra maths symbols
\usepackage{gensymb}

\usepackage{txfonts}
\usepackage{url}
\usepackage[utf8]{inputenc}
\usepackage{subfigure}
\usepackage{float}

\begin{document}

%\begin{frontmatter}

\title{On the rate of core collapse supernovae in the Milky Way}
%\title{More accurate determination of halo environment in the cosmic field}
%\title{An improved algorithm of classification of the cosmic web}

\author{Karolina Rozwadowska$^a$, Francesco Vissani$^a$, Enrico Cappellaro$^b$\\
\small
$^a$ INFN, Laboratori Nazionali del Gran Sasso, Assergi (AQ), Italy  \\
\small and Gran Sasso Science Institute, L'Aquila, Italy\\ \small 
$^b$ INAF, Osservatorio Astronomico di Padova, Padova, Italy}
%\address{Instituto de Astronom\'ia, Universidad Nacional Aut\'onoma de M\'exico, Ciudad Universitaria, 04510, M\'exico.}

\date{}

\maketitle

\begin{abstract}

Several large neutrino telescopes, operating at various sites around the world,
 have as their main objective the first detection of neutrinos emitted by a gravitational collapse in the Milky Way. 
 The success of these observation programs depends on the rate of supernova core collapse in the Milky Way, $R$.
 In this work, standard statistical techniques are used to combine several independent  results. 
\fra{Their consistency is  discussed and the most critical input data are identified.}
The inference on $R$ is further tested and refined 
by including direct information on the occurrence rate of 
gravitational collapse events in the Milky Way and in the Local Group, obtained  
from neutrino telescopes and electromagnetic surveys.
%Events from the Local Group are used as a consistency check on the galactic core collapse rate. }
% {The tensions among the included information are examined 
 %and the role of the most critical inputs is discussed. 
A conservative treatment of the errors yields a combined rate  $R=1.63 \pm 0.46$ (100 yr)$^{-1}$; 
the corresponding time between core collapse supernova events turns out to be 
$T=61_{-14}^{+24}$~yr. 
\fra{The importance to update the analysis of the stellar birthrate method is emphasized.}
\end{abstract}
%
%\begin{keyword}
%supernovas; neutrinos; detectors
%\end{keyword}

%\end{frontmatter}
\parskip1.8ex

\section{{Introduction}}
The rate of core collapse supernovae (CCSN) in the Milky Way, $R$, is a quantity of key importance for the 
current and next generation neutrino telescopes, which includes multi-kiloton detectors such as 
Super- \cite{sk} and Hyper-Kamiokande \cite{hk}, IceCube \cite{ic}, JUNO \cite{jn} and DUNE \cite{dn}; 
in fact, a quantitative evaluation of the supernova rate is relevant also for  multi-wavelength galactic astronomy, 
cosmic ray physics and astrophysics at large.
Various theoretical expectations are available in the literature,
and they are in the range of one-to-few events per century. 
See e.g., $R=1.37\pm 0.74$ (100 yr)$^{-1}$ \cite{cp} for a classical result,
where we updated the value of the Hubble constant \cite{pl} but the error does not include systematic uncertainties, 
and  $R=3.2^{+7.3}_{-2.6}$ (100 yr)$^{-1}$ \cite{bea}, for 
a new determination  based on a SN observability model of the Milky Way, based on 
the small sample of 5 historical galactic SNe (of which only two CCSN) that 
results in a very conservative and conversely not very informative rate.

\fra{
In this work we would like to attempt a more precise evaluation by combining various results that have been presented over the last twenty years in the scientific literature.  We obtain a determination of $R$ with a formal uncertainty of 30\%, which is quite precise, for which the evaluation of $R$ based on the stellar birth rate has a possibly undeservedly high weight. We also discuss the inclusion of the neutron star birth rate, which indicates a higher value in moderate tension with other determinations of the CCSN formation rate. We include in the determination of $R$ the observational results from  
the Milky Way and in the rest of the Local group of galaxies.} 
%A special effort is devoted to a consistent treatment of the uncertainties that, taken at face value from the original references  cannot be directly compared.

\section{A combined rate from the existing literature} \label{brunr}

Our first estimate of the rate of CCSN is based on the combination of the four results obtained by: 
$i)$ counts of massive stars, $ii)$ census of supernova explosions in the cosmos,  
$iii)$ counts of neutron stars and $iv)$ measurement of the galactic chemical enrichment of $^{26}$Al. 
\fra{The combined value is calculated, tested with an independent and {\em very} recent determination (based on SN remnant age determination)
and discussed in Sect.~\ref{sec:combined}.}
Although all these results are based on methods that are well recognised as valid and informative, it is worth noting that none of them relies directly on the  observations of CCSN in the Milky Way. The contribution of more direct - though less informative - 
methods to the global fit of the rate will be examined in the Section~\ref{nexts}. 

\subsection{Counts of massive stars}

 This method is based on the study of slightly more than 400 stars in the Solar neighbourhood to determine the birthrate of stars with mass higher than 10 solar masses~\cite{rd}; this mass limit is adopted to make sure that the stellar catalogue is complete and the method assumes that the minimum mass, leading to the CCSN, is less than this value.
 From this sample the author  derived an estimate of the galactic supernova rate that is quoted as 
  ``probably not less than $\sim 1$ nor more than $\sim 2$ per century'', that one may be tempted to read  as $R=1.5\pm 0.5$ (100 yr)$^{-1}$. 
  \fra{Taken on face value, this result is the most  precise among the available ones,   and for this reason extensive discussion is deserved.}   
  
  We note that, among the possible hidden systematics, a most severe bias is related to the uneven star distribution in the Galaxy. In particular \cite{sc2014} argue that the supernova rate in a region of 1 kpc from the Sun is greater than the galactic mean value by a factor of 5-6. If this is true, it could be risky to extrapolate the local value to the entire Milky Way.
%\fra{But most importantly, it seems incautious to accept the statement $R=1.5\pm 0.5$, first of all because 
%the error is not formally estimated, and  an error of 33\% is much less than the errors of {\em all} the other methods, 
%discussed below, that incidentally are all more recent than this one. 
\fra{In order to proceed conservatively, without discarding arbitrarily this result, 
we adopt as representative the central value indicated by the analysis, but we 
conservatively assume the error to 50\% of the central value.} Therefore, 
defining as usual,
\begin{equation} 
G(R, \bar{R},\delta R) = \frac{\displaystyle \exp \left[ -  \frac{(R-\bar R)^2}{2\ \delta R^2} \right]}{\sqrt{2\pi}\, \delta R} 
  \end{equation}
 we will assume 
 \begin{equation} 
%\begin{aligned}
     \mathcal{L}_{\mbox{\tiny \begin{minipage}[h]{6mm}{\begin{center} stellar\\birthrate \end{center}}\end{minipage}}}(R) = 
  G(R,\ \bar{R}_{\mbox{\tiny sb}},\ \delta R_{\mbox{\tiny sb}})
  \mbox{ with }\; 
  \left\{
  \begin{array}{c}
  \bar{R}_{\mbox{\tiny sb}} =1.5\;  \\[1ex] \delta R_{\mbox{\tiny sb}}=0.75
  \end{array}
  \right.
  %\end{aligned}
  \end{equation}
We refer to this method as {\em stellar birthrate}.    
\fra{The rest of this paper shows that this result has a relevant weight in the combined fit.
%, although with our conservative procedure of increasing the error we have mitigated somewhat its impact. 
 %(Of course, 
 Note that, accepting the nominal error would give an even higher weight to this result in the combined fit; a 
 uniform distribution between 1 and 2 would produce an even stronger effect.}
%These observations alone are a good reason to update and improve this result in the future.

\subsection{Extragalactic SN rates}

A statistical sample of extragalactic CCSN can be used to infer the expected rate 
on the Milky Way assuming that our Galaxy has average properties for its morphologic type 
and luminosity. Ref.~\cite{li} reports  $R=2.30\pm 0.48$ (100 yr)$^{-1}$, that 
updating the Hubble constant \cite{pl} becomes $R=1.95\pm 0.41$ (100 yr)$^{-1}$. 

\fra{This would mean a narrow  
Gaussian distribution $G$, but the attribution of morphologic type for the Milky Way 
 is uncertain and requires to include a systematic error of a multiplicative factor of  $\sim 2$ as stated in~\cite{li}. 
The inclusion of the multiplicative factor of uncertainty is described by a  
log-normal distribution.\footnote{This is the distribution  
$\ell(x,\mu,\alpha) =\exp[ -(\log (x/\bar{x}))^2/(2 \log(\alpha)^2) ]/$ $(\sqrt{2\pi} \log(\alpha)\, x)$
that can be derived by the Gaussian $G(y,\mu,\sigma)$
by changing variable and rewriting the 
average value as $\mu=\log(\bar{x})$; since the values between
$\alpha\bar{x}$ and $\bar{x}/\alpha$ have to correspond to the 1$\sigma$ region,
we set the variance  $\sigma=\log(\alpha)$. In our case we have $\bar{x}=1$ and following \cite{li} 
we will set $\alpha=2$.}
Thus, we consider a multiplicative} coefficient $x$ with distribution
\begin{equation} \label{ellon}
\ell(x) = e^{ -(\log x/\log 2)^2/2}/(\sqrt{2\pi} \log(2) x)
\end{equation}
that has median $x=1$  and obeys 
$\int_{1/2}^2 \ell(x) dx= 68.3\%$. 

The distribution of the true rate $R=x\times R_0$ is obtained changing variable  in the 
the two-dimensional distribution 
 $ G(R_0) \ell(x) dx dR_0 $  and 
 integrating away $x$:  
     \begin{equation}
       \begin{array}{c} \displaystyle
 \mathcal{L}_{\mbox{\tiny \begin{minipage}[h]{8mm}{\begin{center} cosmic\\census \end{center}}\end{minipage}}}(R) = 
 \int_0^\infty   G\!\left(\frac{R}{x},\ \bar{R}_{\mbox{\tiny cens}},\ \delta R_{\mbox{\tiny cens}}\right)\, \ell(x)  \frac{dx}{x}  
 \\[2ex]
   \mbox{ with }\; 
  \left\{
  \begin{array}{c}
  \bar{R}_{\mbox{\tiny cens}} =1.95\;  \\[1ex] \delta R_{\mbox{\tiny cens}}=0.41
  \end{array}
  \right.
  \end{array}
    \end{equation}
 This likelihood corresponds to $R\in [0.93,3.96]$  (100 yr)$^{-1}$
 at 68.3\% CL, \fra{which is much wider than the range from the previous method.}
 We call this method {\em cosmic census} of CCSN.

 \subsection{$^{26}$Al abundance}
 
 This method  models the gamma-ray emission from radioactive $^{26}$Al in the Milky Way.
 Assuming that this emission 
 traces the ongoing nucleosynthesis 
 pollution by CCSN  in the Milky  Way, this method  was used to infer the value 
 $R=1.9\pm 1.1$ (100 yr)$^{-1}$ \cite{diehl}. 
 The uncertainty can be used directly as a Gaussian error: 
 $\mathcal{L}_{\mbox{\tiny Al-26}}(R) = G(R, \bar{R},\delta R) $ with $\bar{R}=1.9$ and $\delta{R}=1.1$.
  The shorthand for this method is simply {\em Al-26.}
Again, note that this range is 
 much wider than the range from the stellar birthrate method.
For  
 a recent study of the role that could be played by $^{26}$Al from young stars see~\cite{refis}.

\subsection{Neutron star birthrate}

  The birthrate of Galactic neutron stars was estimated by \cite{ns} by summing 4 contributions, supposed to be independent: ordinary radio pulsars,  rotating radio transients,
X-ray dim isolated neutron stars,  magnetars. Their rates, in units of objects per century
are respectively: 
$1.6\pm 0.2,$  
$3.2\pm 1.2,$  
$2.1\pm 1.0,$
$0.3\pm 0.3$, obtained by  
averaging the three determinations of Table~1 from \cite{ns} and 
using model NE2001.
We sum the errors linearly and find a conservative result, that we use in Gaussian statistics  with: 
\begin{equation}
\bar R_{\mbox{\tiny NS}}=7.2\mbox{ (100 yr)}^{-1}\ , \ \delta R_{\mbox{\tiny NS}}=2.7\mbox{ (100 yr)}^{-1}\ ,
\end{equation}
This estimate is significantly higher than the estimate derived with other methods.
The question of consistency of this result with the other ones was discussed in the same paper 
\cite{ns}. Using our estimations, if  
we compute
$|\bar{R}_{\mbox{\tiny NS}} - \bar{R}_{\mbox{\tiny Al-26}}|/\sqrt{\delta R_{\mbox{\tiny NS}}^2 + \delta R_{\mbox{\tiny Al-26}}^2}=1.8$   
proceeding as in the appendix of~\cite{pppp}, we would conclude that this result can be attributed to a fluctuation of a 
Gaussian distribution with a chance of 7\%, which is still acceptable.

 \begin{figure*}[t]
\centerline{\resizebox{1\hsize}{!}{\includegraphics{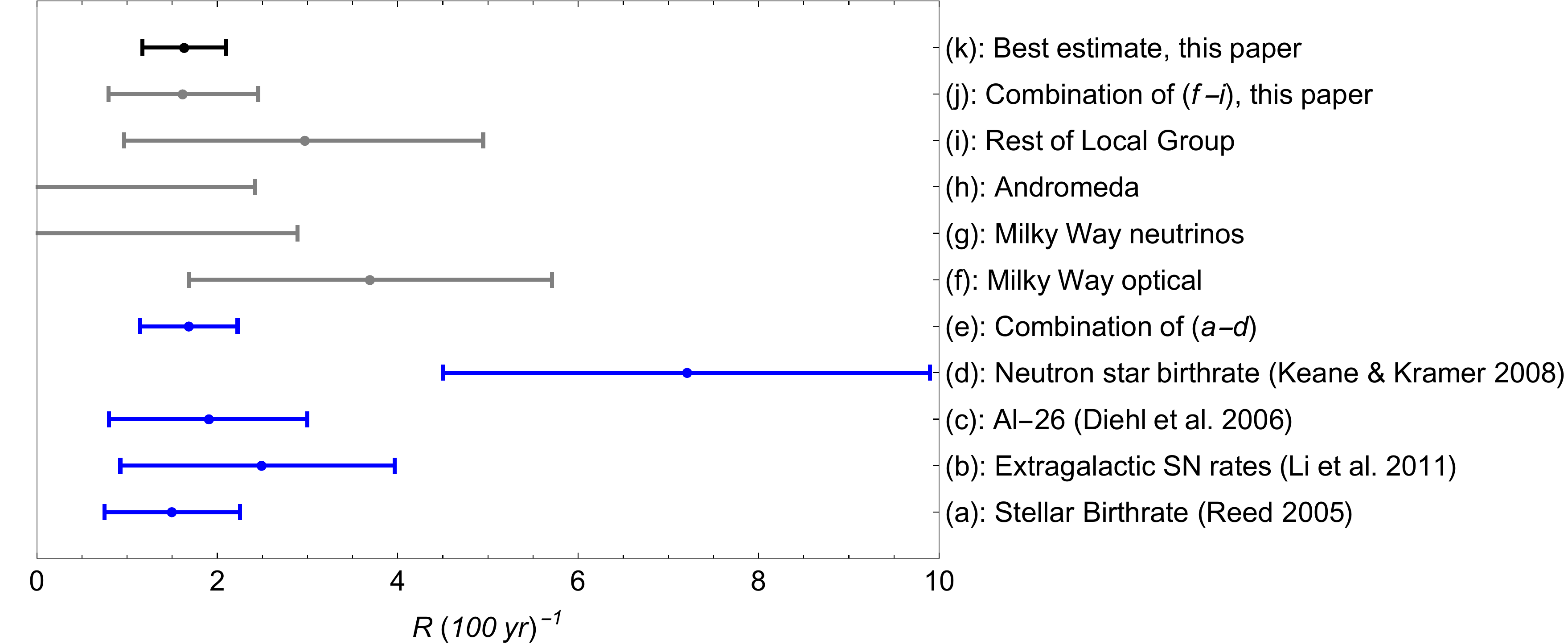}}}
\caption{\small\em CCSN rate $R$ in the Milky Way: from the existing literature (blue), computed from direct information from the Local Group (gray) and 
full result (black) of Eq.~(\ref{eq:full}).
\label{fig1}}
\end{figure*}

\subsection{Model for SNR ages \label{cagoja}} 

\fra{Recently \cite{l1,l2}  used a theoretical model to reconstruct the properties of  58 supernova remnants (SNR). The fraction of     CCSN in this sample is $f_{0}=0.78\pm 0.08$.\footnote{This is based on the sub-sets of classified SNRs: $21-30$ CC and $6-9$ Ia.} 
The estimation of the age of the SNRs allows to calculate their production rate $r$; this, together with an estimation of the incompleteness factor of the sample $\eta$, yields the total rate of supernovae formation  (CCSN+Ia) 
according to the formula $\mathcal{R}=r\times \eta$. 
However~\cite{l2} argues that the original theoretical model does not properly 
describe the case of supernovas  embedded in a medium with the density profile of a stellar wind.
After proper analysis of the wind dominated SNR, 
 \cite{l2}  derived an effective  correction factor $\delta$ which is estimated to be in the range $\delta \in [0.19, 0.41]$, and can be introduced in the formula above as follows: $\mathcal{R}=r\times (1+\delta)\times \eta$. 
 If we want to limit to the rate of CCSN, we can write $R=r \times (f_0 +\delta f_1)\times \eta$,  where $f_1$ is the fraction of CCSN subject to the correction. From Table~4 of~\cite{l2}, one finds that  $f_1\sim 0.5$ with an uncertainty that is unlikely to be less than the one of $f_0$; thus we use the conservative range $f_1\in [0.4,0.6]$.
%The shorthand for this method is {\em SNR model.}
Coming to numerical estimates, we find  a median of 
$R=1.7$ (100 yr)$^{-1}$. The 68.3\% range of the coefficient is in the interval $[0.8, 3.5]$,  
where the main uncertainty is the  factor of 2 on $\eta$ \cite{l2}. 
This is a very interesting result, and  possibly along similar lines there will be future progress. 
For the time being, however, we prefer to use this result as a test  rather than including it with the other determinations, for a number of reasons:  
1)~The value of $r$ that we obtain is\footnote{This was checked  in several ways: by considering the number of SNR $N(T)$ whose age is less than a time $T$, and evaluating $r=N(T)/T$ (the effect of the uncertainties on the age was also estimated); fitting the cumulant distribution with a linear function $N(T)=\kappa + r T$, with $\kappa=0$ as in \cite{l1} or letting $\kappa$ free to fluctuate as recommended in \cite{l2}. The value that we find does not change significantly and the range  is consistent with Poisson fluctuations.} 
$(0.37\pm 0.05)$  (100 yr)$^{-1}$, which is smaller than the value  $r=1/230\mbox{ yr}\sim 0.43$/century cited in \cite{l2}.
2)~The correction $\delta$ is large, and adopting the average value of the interval might introduce a bias in the analysis. Note that 
the fraction of CCSN decreases after including $\delta$, becoming $(f_0 +\delta f_1)/(1+\delta)=0.72_{-0.07}^{+0.08}$.
3)~The works \cite{l1,l2} are primarily aimed at modelling the 58 SNR, and in our understanding the determination of total rate of SNR is mostly a consistency check. In conclusion, we regard the numbers obtained above as preliminary and 
suggest that an independent study of SNR, specially dedicated to determining the CCSN rate, should be attempted.}

\subsection{Result of the statistical combination} \label{sec:combined}
 
The four results listed above employ different methods with also different error estimates.
In fact, the methods do not report consistent systematic errors, especially the {\em stellar birthrate} uses several assumptions, not indicating the exact error. 
Anyway, these published results are consistent and can be combined to obtain a more precise inference. 
Multiplying the four likelihoods, we find
    \begin{equation}
  R_{\mbox{\tiny combined}}=1.79\pm 0.55\ (100\mbox{ yr})^{-1}
  \label{eq:comb}
  \end{equation}
  \fra{which is  very close to the value we derived from the SNR model as described in Sect.~\ref{cagoja}.}
  Unless stated otherwise, we will always 
   quote as a rule the {\em average values} for $R$, that do not differ by much from the mode for all cases. 
\fra{We note that the \textit{stellar birthrate} input remains the most important 
individual result in this combined fit, despite our conservative treatment of the errors.
This is quite evident from the visual comparison allowed by Fig.~\ref{fig1}, lower part. 
Therefore, it is important to add independent  information; this is the goal of the next section.}

\begin{table*}[t]
\caption{\em\small Observed CCSN in the Local Group.}
\label{tab:rates}
\centering
{\small
\begin{tabular}{ccccc}
\hline \hline
number of  & observation &  location & observational & assumptions, \\
CCSN & time [yr] & approach   &  method & see \ref{app:a} and \cite{bea} \\  \hline
0 & 40 & Milky Way & neutrino telescopes & direct \\
{0} &  {100} & Andromeda & optical surveys &  $f \in [0.25, 0.75]$ \\
{1} & {100} & rest of Local Group &  optical surveys & $f \in [0.40, 1.05]$ \\
{2} & {1300} & Milky Way &  historical & $\varepsilon=1/16$ \\
\hline
\end{tabular} }
\end{table*}

\section{Core collapse supernovae in the Local Group} \label{nexts}

\subsection{Milky Way}
The historical records of astronomical observations of the Milky Way date back to 4th century BC, with  more reliable records starting possibly from 8th century \cite{china}. In the last 40 years the neutrino telescopes joined the CCSN observational campaign. We use the CCSN observed by means of  visual and neutrino astronomy (2 and 0 CCSN, respectively) to directly constrain the galactic rate of CCSN.

\subsubsection{Ancient astronomy}
Chinese astronomy flourished particularly 
after the astronomer Yi Xing, namely, after 8th century (Tang dynasty; see \cite{china}): therefore we assume an essentially 
complete astronomical coverage for $T=13$ centuries. 
In this period, 2 CCSN have been seen in the Milky Way:\footnote{Chinese historical records exist since the 4th century BC, 
and  1  (perhaps 2) CCSN have been seen in this period:  SN 393 (and SN 185). 
We do not use these data as the uncertainty is considerable.} 
SN 1054 and SN 1181 \fra{\cite{cshs}, which are the very same  CCSN used in the analysis of \cite{bea}.}
%The occurrence of CCSN is a Poisson process,  which implies that the most plausible number of visible CCSN accessible to optical astronomy is just 2.  
The occurrence of CCSN events is a Poisson process, which implies that the most plausible number of visible CCSN accessible to optical astronomy is precisely two.  

We know that,  most CCSN in the Galaxy are hidden in the visible light due to dust obscuration.\footnote{\fra{A clear demonstration of the existence of such objects is given by Cas A, which is a supernova remnant associated with a X-ray point-like source 
that is about 300 years old, but was not seen at the time.}}
The fraction of  visible CCSN according to the calculations of 
 \cite{bea}   is $\varepsilon=1/16$; in fact, they 
 estimate that the best-fit in the last millennium is 3.2 CCSN/(100 yr). 
The error  is not estimated, however, the comparison of the value of $\varepsilon$ with that obtained by 
another method of calculation \cite{bea} suggests that it is small and does not give an effect as 
important as the one of statistical fluctuations; therefore, we neglect it.

 For these reasons 
 we will adopt the following likelihood over~$R$
 for the historical galactic CCSN
\begin{equation}
{\mathcal{P}}_{\mbox{\tiny \begin{minipage}[h]{9mm}{\begin{center} Milky~Way \\astronomy \end{center}} \end{minipage}}} = \frac{\mu^2\ e^{-\mu}}{2}    \mbox{ with }\mu= \varepsilon\ R\ T
\end{equation}

\subsubsection{Neutrino telescopes}
 
Thanks to the neutrino detectors  Artemovsk, Baksan, Kamiokande-II, LVD, Super-Kamio\-kande {\em etc.}
we are sure that we have not seen neutrinos from 
CCSN events in the Milky Way in the last $\sim$ 40 years. 
The information concerning neutrinos applies directly  to CCSN  being 
therefore particularly important; see \cite{sknu,lvd}  for further discussion.

The time of observation of the Milky Way is 
 $T=39.7$ yr.  In order to obtain  this value, we consider the dates of beginning of data taking in Artemovsk (0.1 kt of scintillator \cite{ar})
 $T_1=1977+11/12$, of the beginning of data taking in Baksan (0.2 kt of scintillator \cite{ba})  
 $T_2=1980+6/12$, of the beginning of data taking in Kamiokande-II (2.14 kt of water  \cite{ka}) 
 $T_3=1987$, and the present date  
  $T_4=2020+2/12$. Then we calculate 
  \begin{equation}T=\varepsilon_{\mbox{\tiny A}} (T_2-T_1) +   \varepsilon_{\mbox{\tiny B}} (T_3-T_2) + \varepsilon_{\mbox{\tiny C}} (T_4-T_3)\end{equation}
  where we estimate the efficiency to detect a CCSN as $\varepsilon_{\mbox{\tiny A}} =0.5$ 
 in    Artemovsk  (in view of its small mass), 
as $\varepsilon_{\mbox{\tiny B}} =32.1/(2018 - 1980.5)=0.86$   in Baksan \cite{ba}
and, 
  after the beginning of 
  Kamiokande-II operation, we set $\varepsilon_{\mbox{\tiny C}}=0.99$;
  the third term is by far the largest contribution.

      We adopt again a Poisson likelihood for $R$, that describes the 
   lack of neutrino observation of the CCSN within the Milky Way, 
\begin{equation}
{\mathcal{P}}_{\mbox{\tiny \begin{minipage}[h]{9mm}{\begin{center} Milky~Way \\[-0.2ex]neutrino \end{center}} \end{minipage}}} =   e^{-\mu}     \mbox{ with }\mu=   R\ T
\end{equation}

\subsection{External check: CCSN in Andromeda and in the rest of the Local Group } \label{pech}
We can perform an external check on the consistency of the above estimate by examining the record of CCSN events in the Local Group, outside the Galaxy.

The Local Group of galaxies includes more than 50 galaxies. The Galaxy and M31 (Andromeda) are the largest; all other are small or very small and do not add much to the integrated luminosity of the Local Group. However, in irregular galaxies, such as many of those in the rest of the Local Group, 
 the  probability  (per given mass) of having a CCSN is {\em a priori}  larger. 

The galaxies of the Local Group have been monitored for optical transients 
since 1885  at least when a type Ia supernova exploded in M31. Actually, the first SN search was started by Fritz Zwicky only in 1935. 
More recently, the core collapse  SN1987A in the LMC was discovered.
In this manner, we have learned that in about one century only 1 CC SN was found in the Local Group.

The corresponding likelihoods are
\begin{equation}
\mathcal{P}=\mu^n \exp(-\mu) \mbox{ with } \mu = fR \times T
\end{equation} where for $T$ we take conservatively 100 yr. 

The rates in Andromeda and the rest of the Local Group were expressed as $f R$, where we remember that $R$ is the rate in the Milky Way: therefore, 
$f$ describes the relative rate of CCSN compared to that in the Milky Way and the possibility that some of the events were not observed, due to, for example, the absorption of the emitted light.  In order to proceed conservatively,   
we do not  fix the value of $f$ 
 as was done in \cite{lila1}, but 
we  average the Poisson likelihoods over $f$, replacing  $\mathcal{P}$ with  
\begin{equation}
\langle {\mathcal{P}} \rangle=
\frac{1}{f_2-f_1} \displaystyle \int_{f_1}^{f_2} \mathcal{P}(f\, R T)\ df
\end{equation} 
that have simple analytical expressions. 
Based on the observed samples of the supernova remnants 
in the Local Group we assume $f \in [0.25, 0.75]$ for M31, and $f \in [0.40, 1.05]$ for the rest of the Local Group,
for the reasons discussed in details in \ref{app:a}.

\subsection{Combined result from core collapse supernovae in the Local Group}

%
%\begin{table}
%\caption{Observed CCSN in the Local Group.}
%\label{tab:rates}
%\centering
%\begin{tabular}{rrr}
%\hline \hline
%number of  & observation &  location and \\
%CCSN & time [yr] & remarks \\  \hline
%0 & $\sim$ 40 & MW, neutrinos \\
%\multirow{2}{*}{0} & \multirow{2}{*}{100} & Andromeda, optical surveys;\\
%&& $f \in [0.25, 0.75]$ \\
%\multirow{2}{*}{1} &\multirow{2}{*}{100} & rest of LG, optical surveys;\\ && $f \in [0.40, 1.05]$ \\
%\multirow{2}{*}{2} & \multirow{2}{*}{1300} & MW, historical;\\
%&& $\varepsilon=1/16$ \\
%\hline
%\end{tabular} 
%\end{table}
%

\begin{figure}[t]
\centerline{\resizebox{0.9 \hsize}{!}{\includegraphics{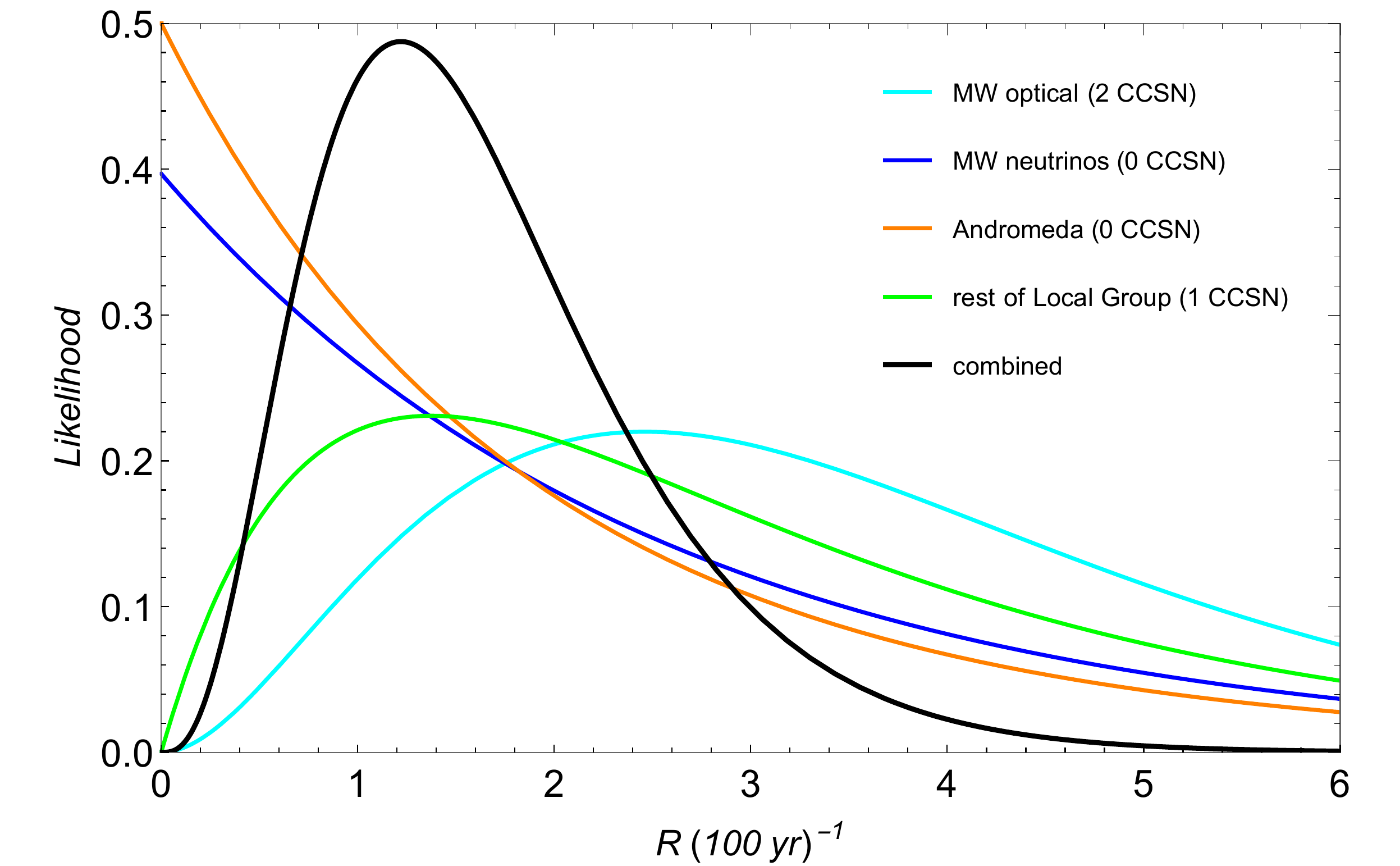}}}
\caption{\em\small Likelihoods of the CCSN rate $R$ in the Milky Way from the historical evidence of CCSN in the Galaxy and Local Group.  
For comparison, we show also the likelihood discussed in Sect.~\ref{sec:combined}, indicated with the label `combined'.
\label{fig:hist}}
\end{figure}

The four likelihoods discussed in this section are shown in Fig.~\ref{fig:hist}.
By combining them, 
\begin{equation}
 \mathcal{L}_{\mbox{\tiny Local CCSN}}  = 
 {\mathcal{P}}_{\mbox{\tiny \begin{minipage}[h]{7mm}{\begin{center} Milky~Way \\astronomy\end{center}} \end{minipage}}} 
 \times   {\mathcal{P}}_{\mbox{\tiny \begin{minipage}[h]{7mm}{\begin{center} Milky~Way \\neutrino \end{center}} \end{minipage}}} 
 \times  \langle{\mathcal{P}}_{\mbox{\tiny  \begin{minipage}[h]{3mm}{M31}\end{minipage}}}   \rangle
 \times   \langle {\mathcal{P}}_{\mbox{\tiny \begin{minipage}[h]{9mm}{\begin{center} rest of Local\\[-0.4ex] group \end{center}}\end{minipage}}}   \rangle
\end{equation}
we find the distribution shown in black in Fig.~\ref{fig:hist}. Let us discuss this distribution. 
Its mode,  median and average are slightly different among them:
\begin{equation} \label{eq:local}
R_{\mbox{\tiny Local CCSN}}=1.22, \ 1.48,\ 1.62\ (100\mbox{ yr})^{-1}
\end{equation}
The 68.3\% interval, built using two equal tails at the left and right of the distribution, turns out to be
\begin{equation}
R_{\mbox{\tiny Local CCSN}}=(0.84,2.41)\ (100\mbox{ yr})^{-1}
\end{equation}
while if this interval is built integrating the likelihood, when it is above  
a certain value (which resembles a bit the Gaussian construction of confidence levels) this is 
\begin{equation}
R_{\mbox{\tiny Local CCSN}}=(0.66,2.04)\ (100\mbox{ yr})^{-1}
\end{equation}
\fra{The fact that these two interval do not 
coincide exactly, just as  the mode, the median and the average do, 
shows that the combined likelihood is slightly non Gaussian.
However, Fig.~\ref{fig:hist} does not 
indicate any critical inconsistency of these input data among them, or with those discussed in 
Sect.~\ref{sec:combined}. This is further illustrated visually in the upper part of 
Fig.~\ref{fig1}.
For this reason, we proceed and combine the entire dataset.}

\section{Full result}
When we proceed and use the full information - namely, the one from the combined rate, together with the  one from 
neutron star birthrate and the direct information on the rate of CCSN in the Local Group,  
the resulting likelihood  becomes
 \begin{equation}
 \mathcal{L}_{\mbox{\tiny full}}  \propto  \mathcal{L}_{\mbox{\tiny combined}} 
 % \times  \mathcal{L}_{\mbox{\tiny NS birthrate}} 
% \exp\left[  - \frac{(R - \bar{R}_{\mbox{\tiny NS}}  )^2}{2\, \delta  R^2_{\mbox{\tiny NS}}  }\right] 
 \times   \mathcal{L}_{\mbox{\tiny Local CCSN}}
 \end{equation}
This implies the following range for the CCSN rate in the Milky Way
  \begin{equation}
    R_{\mbox{\tiny full}}=1.63 \pm 0.46\ (100\mbox{ yr})^{-1} 	
    \label{eq:full}
\end{equation}
The likelihood resulting from combination of several Poissonian and Gaussian distributions results in a quasi-Gaussian distribution, with median 1.61 (100 yr)$^{-1}$ and mode 1.57 (100 yr)$^{-1}$.
The  $ \chi^2/(N-1)$ test\footnote{The $\chi^2$ can be evaluated directly  from the average value $\bar{x}$ and  the likelihoods $\mathcal{L}_i$ by the formula  $\chi^2= -2 \sum _{i=1}^N \log[ \mathcal{L}_i( \bar{x} ) / \mathcal{L}_i( \bar{x_i}  ) ]$. When the likelihoods are almost Gaussian as in our case the results are in good agreement.}
 of our sample suggests that the input values are consistent. Taking all four results from Sect.~\ref{brunr}, contributing to the \textit{combined rate} and combination of historical CCSN in Milky Way and Local Group from Eq.~(\ref{eq:local}), $\chi^2/(N-1) = 1.01$ with $N=5$. The highest contribution comes from the neutron star birthrate result of \cite{ns}.

The value $R_{\mbox{\tiny full}}$  is quite similar to  the result from Eq.~(\ref{eq:comb}), based only on results of the previous 
scientific literature. 
Eq.~(\ref{eq:full}) is the best value that we can offer in view of present information,
 and for convenience we will call it the {\em best estimate}.
Comparison of the cited values with our result is shown in Fig.~\ref{fig1}. Let us comment on this result.
 The  increase of the expected rate,  due to the inclusion of the neutron star birthrate in the analysis,   
 is  almost exactly compensated by the decrease  due to the inclusion of the 
 direct information from CCSN events in the Local Group. 
 In view of the Poisson character of the CCSN likelihood, and due to the very meagre dataset 
 (i.e., 2 CCSN in Milky Way over 13 centuries and 1 CCSN in Large Magellanic Cloud) 
 the last distribution will need to be updated as soon as new events  are observed.
  
  The {\em stellar birthrate} results of \cite{rd}, which we treat in a conservative way by increasing the error, is still the input with the smallest
  error. 
    \fra{As a matter of fact, we have not found any valid reason to discard this result, neither on a physical basis nor in comparison with the other results: the combined result is almost indistinguishable to the  1$\sigma$ range $1<R<2$, that is 
  $R=1.5 \pm 0.5\ (100\mbox{ yr})^{-1}$. 
We believe that, while these considerations do not motivate particular doubts towards the combined result, they offer good reasons to undertake an updated analysis and more accurate assessment of the results on CCSN from the stellar birth rate method. This might be particularly desirable considering the results  of the GAIA \cite{gaia} mission.}

\section{Summary and discussion}
 
 In summary, we have shown that various determinations of the rate of CCSN in the Milky Way, 
 when combined, produce a likelihood that is distributed in a quasi-Gaussian way. 
 This likelihood indicates a rate of $R=1.63 \pm 0.46$ CCSN per century, see Eq.~(\ref{eq:full}). 
 Note that this result is consistent but more precise than
 the values of  \cite{cp} or of \cite{bea} cited in the introduction,    
and therefore quite informative, being 
in particular at odds with 
 the most optimistic (higher) CCSN rates. 
    \fra{This conclusions shows that it is urgent to assess and clarify  current expectations based on 
neutron star birthrate \cite{ns}, which is the one among the ones we have considered 
that indicates the largest value of the rate.
The most crucial individual input however remains the {\em stellar birthrate} method \cite{rd}.}

 It is also interesting to discuss the value for the time of CCSN events in the Milky Way, $T=1/R$.
 The corresponding likelihood has a considerable skewness, 
 and consequently the expected range is quite asymmetric. 
 We report the time of CCSN as a 68.3\% CI around the median
\begin{equation}
T_{\mbox{\tiny full}}=61_{-14}^{+24}\mbox{ yr}
\end{equation}
while the most plausible value (mode) is at 52~yr.
  By way of comparison, in previous authoritative works the value of $T$ remains 
  firmly fixed around 50 yr: for example, the monograph of Ginzburg and Syrovatskii ii 
 \cite{gs64} 
 uses $T=$50 yr in Eq.~(11.9), and the review work of Tammann, Loeffler and Schroeder 
 \cite{t94}, that  summarizes most of the information available in the last century,  
 quotes $T=40/0.85=47$ yr already in its abstract.
 
A neutrino telescope that runs for a time $t$ has the Poissonian chance $P(0)=e^{-R t}$  that 
no galactic supernova will occur during its running time. When the uncertainties  on the rate $R$ 
are described by a Gaussian distribution restricted to the physical region $R>0$, the chance of observing at least one event, $P(>0)=1-P(0)$, is given by
\begin{equation}
\displaystyle
P(>0)=1-e^{- R t}  
 \cdot \frac{{ \left[ 1+\mbox{erf}\left( \frac{R-\sigma^2 t}{\sqrt{2} \sigma} \right) \right] \cdot e^\frac{\sigma^2 t^2}{2}}  }{1+\mbox{erf}\left( \frac{R}{\sqrt{2} \sigma} \right)}
\end{equation}
where `erf' indicates the Gauss error function. 
With the best fit values obtained above for the rate $R$ and its variance $\sigma$, Eq.~(\ref{eq:full}), we find that the chances of seeing at least one event  are 
$P(>0)= 7.9\%$, 15.0\% and 27.4\% and 53.7\% when the time of observation is $t=5, 10, 20$ and 50 years respectively. This consideration illustrates 
quantitatively the importance of disposing of  neutrino telescopes that are stable and capable to operate for a rather long time.

\subsection*{Acknowledgements}
This work was partially supported by the research grant 2017W4HA7S ``NAT-NET: Neutrino and Astroparticle Theory Network'' under the program PRIN 2017 funded by the Italian Ministero dell'Istruzione, dell'Universit\`a e della Ricerca. % (MIUR). 
%All rights reserved. 
We thank M.L.~Costantini, W.~Fulgione, A.~Gallo Rosso, C.~Mascaretti and C.~Volpe for precious discussions.

\newpage

\begin{appendix}

\parskip1ex

\section{Core collapse supernovae in the Local Group}\label{app:a}

\begin{table}[t]
\centering
 \caption{\em\small Number of SNR and fraction of CCSN in the Local Group galaxies.
 The observational methods are in the order from most to least significant surveys.
 } \label{tab:appA}
\begin{tabular}{ccccc}
\hline \hline
\small Galaxy &\small  SNR  & \small CCSN & \small Observational  & \small Ref.  \\ 
  &\small  sample & \small fraction & \small methods &  \\ 
 \hline 
 MW & 294  & \ldots &  \small optical, radio, X-ray & \cite{Green:mw}\\  
 M31 & 156 & 0.75  & \small  optical, X-ray, radio& \cite{Lee:m31} \\  
 M33 & 155 & 0.85 &  \small  optical, radio, X-ray &\cite{White:m33,Lee:m33}
 \\  
 LMC & 59 & 0.52-0.59&  \small  X-ray, optical, radio &\cite{Maggi:lmc}   \\  
 SMC & 21 & 0.73-0.84 &  \small  X-ray, optical, radio & \cite{Maggi:smc} \\ 
 \hline 
 \end{tabular}
 \end{table}

In the main text, we introduce the  relative rate of CCSN  
in M31 in comparison with the Milky Way and the same for 
the rest of the Local Group (mostly M33, SMC, LMC). 
These relative rates, indicated by $f$, 
are based on the number of 
observed supernova remnants (SNRs). 

The information that we use is summarized in Tab~\ref{tab:appA}.
The galactic sample contains almost 300 SNRs listed in \cite{Green:mw} catalogue. 
The studies by  \cite{Lee:m31} of M31 report a sample of 156 SNRs, 
while  \cite{White:m33} identify 155 SNRs with multiwavelength coverage in M33. 
Surveys of  Magellanic Clouds produced lists of 21 \cite{Maggi:smc} 
and 59 \cite{Maggi:lmc} SNRs in SMC and LMC, respectively. 
These samples are most complete, thanks to the favourable nearby position of Magellanic Clouds, 
with small absorption in the line of sight. 
Milky Way, M31 and M33 samples are likely to be incomplete, 
limited by the sensitivity and efficiency of the observations, 
confusion with other sources and absorption in the line of sight. 

The exact classification of SNRs to the Ia-type or to the CCSN-type is difficult, 
we assume that the fraction of core collapse SNRs contribution to the entire sample is similar for the galaxies in the Local Group 
and lies in the range $\frac{n_{\text{CC}}}{(n_{\text{CC}}+n_{\text{Ia}})} \in [0.5, 0.85] $ \cite{Maggi:lmc, Maggi:smc, Lee:m33, Lee:m31}.

Accounting for the observed SNRs and the possible SNR candidates reported in the literature
  \cite{Green:mw, Maggi:lmc, Maggi:smc, White:m33,  Lee:m33, Lee:m31, Long:m33}, 
  we obtain our best estimates $f \in [0.25, 0.75]$ for M31 and $f \in [0.40, 1.05]$ for the rest of the Local Group. 
  
  Thus, the whole 
 Local Group %has $f=2.02\pm 0.21$, that 
 is expected to have about one CCSN in about 30 years,
i.e., %a rate of occurrence of CCSN 
roughly twice than the Milky Way alone.

  The estimates of $f$ agree with simple comparison of the galaxy masses and types together with probability of missing observation of the events.
The mass of M31 is similar to that of Milky Way,  but correction for the high inclination angle of M31 and therefore probable absorption in the galactic disk is needed. In the rest of the Local Group the main SN contribution is expected from:  relatively small M33;  Magellanic Clouds with  small masses compared to the Milky Way, but higher supernovae rate due to irregular structure and high probability of observation from Earth taking into account close distance and good visibility.
    
    \end{appendix}

%\footnotesize
%\begin{twocolumn}
% \bibliographystyle{ieeetr}

\newpage
\small
  \bibliography{ref2}
%\end{twocolumn}

\end{document}